\documentstyle[aps,prl,multicol,enumerate,amstex,amssymb,epsfig]{revtex}
\begin{document} 

\draft 
\flushbottom

\title{Universal Conductance Distributions in the Crossover between
Diffusive and Localization Regimes}

\author{A. Garc\'{\i}a-Mart\'{\i}n and
J.J. S\'{a}enz} 
 
\address{Departamento de F\'{\i}sica de la Materia Condensada 
and Instituto de Ciencia de Materiales ``Nicol\'{a}s Cabrera'', \\
Universidad Aut\'{o}noma de Madrid, E-28049 Madrid, Spain.}

\date{\today} 
\maketitle 

\begin{abstract}   
The full distribution of the conductance $P(G)$ in quasi-one-dimensional 
wires with rough surfaces is
analyzed  from the diffusive to the localization regime.
In the crossover region, where 
the statistics is dominated by only one or
two eigenchannels, the numerically obtained $P(G)$
is found to be independent of the details of the system with the average
conductance $\langle G \rangle$ as the only scaling parameter.
For $ \langle G \rangle < e^2/h$,
 $P(G)$ is given by an essentially ``one-sided'' log-normal distribution. 
In contrast, for $e^2/h < \langle G \rangle \le 2e^2/h$, the shape of
$P(G)$   remarkable agrees with  those predicted by  random matrix theory
for two fluctuating transmission eigenchannels.
\\ 

\end{abstract} 
\begin{multicols}{2}

The  large conductance fluctuations occurring in mesoscopic
samples have attracted large attention during the last
decade\cite{Ehrenreich_Altshuler,Bee}. 
The magnitude of the fluctuations ($\sim
e^2/h$) is almost independent of the mean value of the 
conductance $\langle G \rangle$ and the
system size (`universal conductance fluctuations')\cite{UCF}.
This holds for wires or waveguides in the diffusive regime 
where the length of the wire $L$ is much larger than the mean free path
$\ell$ but still much smaller than the localization length $\xi$.
As the length of wire $L$ approaches the localization length $\xi$,
the  conductance fluctuations become of the same order as the averaged
conductance. The averaged values are then not enough to describe the
transport properties  and the  knowledge of the
conductance probability distribution is of primary
interest. 

While for a single-mode wire, where $\xi \approx \ell$, the  
 probability distribution $P(G)$ can be computed for any system
length,
it is remarkably difficult to extend this result to the N-mode case
\cite{Ehrenreich_Altshuler,Bee}.
Based on 
numerical simulations as well as on 
perturbation theory\cite{Ehrenreich_Altshuler,Bee,Pichard},
$P(G)$ is expected to evolve from a Gaussian (deep in the diffusive regime) 
to a lognormal distribution (deep in
the localization regime). However, the behavior of $P(G)$ 
in the crossover between diffusive and
localized regimes is not well known. 
Recently,  both experimental\cite{Cobden} and numerical 
\cite{Cho,Plerou,Ruhl,Wang,Jovanovic}
studies have
reported that the conductance distribution at an integer quantum hall
transition presents a striking shape in the form of a broad
distribution. Related to that, recent numerical
studies suggested a universal distribution at the metal-insulator
transition\cite{Markos}. Based on  the Dorokhov-Mello-Pereyra-Kumar
(DMPK) equation \cite{DMPK} 
for quasi-1D wires in  absence
of time reversal symmetry
it has been found\cite{Muttalib} that  the crossover region between metallic
and insulating regimes is highly non-trivial, and shows 
{\em one-sided log-normal} distribution for the conductance at the
transition. 
It was suggested \cite{Ruhl,Muttalib} that these distributions might not
be neither a private issue of the quantum Hall transition nor  a product
of the removal of the time reversal symmetry due to the presence of
magnetic fields. Moreover, numerical calculations\cite{Ruhl} in 2D and 3D
suggest that the form of $P(G)$ at the critical point is independent of the
dimensionality of the system and of the model.  

Here we present the results of extensive numerical calculations of the
conductance distribution $P(G)$ and transport eigenchannels  all the way
from the diffusive to the localization regime, paying special attention to
the regime crossover. We show that close to the crossover regime, where the averaged
conductance $\langle G\rangle$ is between 1/2 and 1 (in units of
$2e^2/h$),  the general characteristics of $P(G)$ are dominated  just by
two  fluctuating eigenchannels. The diffusion-localization transition is
fully characterized by a transition from two to only one eigenchannel,
occurring at a critical value $\langle G\rangle_c \sim 1/2$. 
Right at  the critical value $\langle G\rangle_c$, 
the distribution is almost perfectly flat and almost identical to
that obtained for 2D systems in the quantum Hall regime\cite{Wang}.
In the insulating regime
($\langle G\rangle < \langle G\rangle_c $) our 
results are consistent with the one-side log-normal distributions
predicted by the DMPK equation \cite{Muttalib}. 

\begin{figure}
 
 
\narrowtext 
\begin{center} 
\epsfig{file=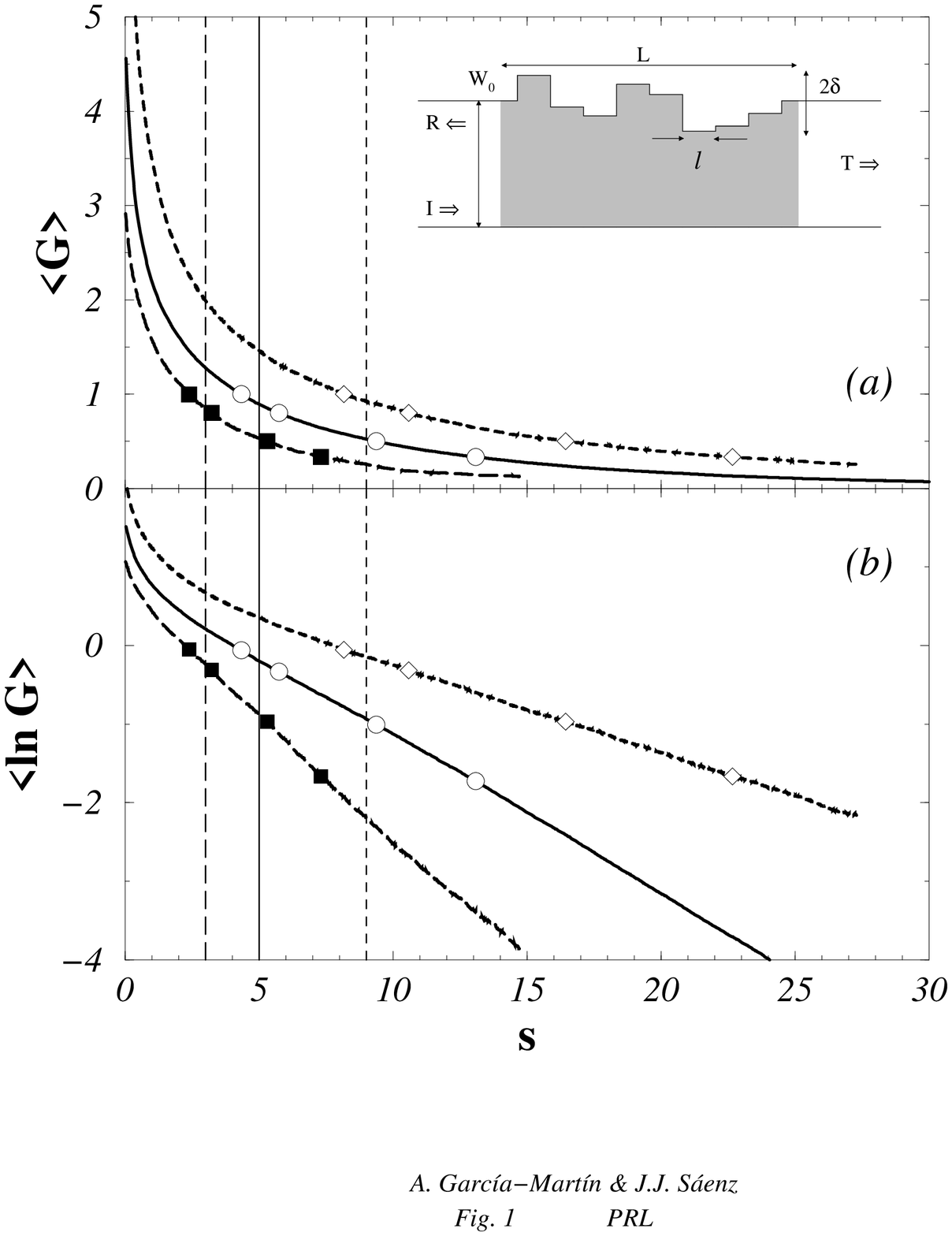,width=8.5cm,clip=}  
\caption{Averaged conductance $\langle G\rangle$ (a)
and averaged logarithm  of the conductance $\langle \ln G\rangle$ (b) as a
function of the length of the disordered part of the wire for 
$W_0/\lambda=4.9$, i.e. 9 modes in the clean part (dashed line), 
$W_0/\lambda=2.6$, i.e. 5 modes (solid line)  and 
$W_0/\lambda=1.8$, i.e. 3 modes (long-dashed line).  Vertical lines
indicate the localization length for each case.
Symbols are the
points  at which the conductance distributions 
have been calculated.
Inset: Schematic view of the system under consideration.} 
\label{System} 
\end{center} 

\end{figure}

Most of the previous work on transport properties of disordered wires has
been focused on the influence of multiple scattering with bulk defects. 
In this work we are going to deal with surface corrugated wires (SCW) 
as a model
system. The interest on these systems is twofold. On 
 one hand,
 for very thin wires, the main source of 
multiple scattering comes only from the
surface roughness.  On the other hand, in absence of bulk defects, 
the transport through SCW,
like in the DMPK approach,  does not contain the effects of wave function
correlations in the transverse direction. 
Many statistical  properties of transport 
through SCW \cite{APL,PRL1,PRL2,SanchezGil} 
have been found to be 
in good agreement with those predicted by 
random matrix theory (RMT)\cite{Bee} for 
bulk defects. In particular, the analytical distributions of 
transmission coefficients \cite{Lagen} obtained from DMPK equation are nicely 
reproduced in numerical simulations of SCW \cite{PRL1}. 
Our model system consists of a surface corrugated
two-dimensional (2D) wire 
(see inset in Fig. \ref{System}).
The corrugated part of the wire, 
of total length
$L$ and perfectly reflecting walls, is composed of  slices of length
$l$. The width of each
slice has random values uniformly distributed between $W_0-\delta$ and
$W_0+\delta$ about a mean value $W_0$. 
The transmission matrix $t$
is exactly calculated for each realization by solving 
the 2D wave
equation by mode matching at each slice, together with a generalized
scattering matrix technique\cite{APL,WeissJosanMetodo}. 
The conductance is given by $G = \text{trace}\{tt^{\dagger}\}$. To obtain 
the mean values ($\langle \dots \rangle$) and the probability
distributions we have used 10000 independent
realizations of the disordered guide. 


We have performed
calculations for three different values of $W_0/\lambda$:
4.9, 2.6 and 1.8, allowing $N=$ 9, 5 and 3 propagating 
modes respectively ($\lambda$ would be the Fermi wavelength for 
electron transport or the wavelength of the diffusive incoming waves in the
case of transport of classical  waves).
We have considered  $l/\delta = 3/2$ and different ratios
$W_0/\delta= 7$ (for $N=5,3$) and 
$W_0/\delta= 13.25$ (for $N=9$).
The behavior of different mean values of transport coefficients as a
function of the length $L$ has been discussed previously in detail
\cite{APL,PRL1}. In the diffusive regime, the averaged resistance 
follows a typical ohmic behavior 
($\langle 1/G\rangle \approx 1/N +  L/\xi$)
while, in the localization regime, $\langle\ln G\rangle \approx -L/\xi$. 
In Figure 1 we plot our results  both for $\langle G \rangle$ 
(Fig. 1a) and for $\langle\ln G\rangle$ 
(Fig. 1b) versus $s \equiv N L/\xi$ for 
$N=9$ (dashed line), $N=5$ (solid line)
and $N=3$ (long-dashed line). The corresponding localization lengths are
$\xi=187 W_0$ ($N=9$),  $\xi=34.4 W_0$ ($N=5$) and
$\xi=51 W_0$ ($N=3$).

\begin{figure}
 
 
\narrowtext 
\begin{center} 
\epsfig{file=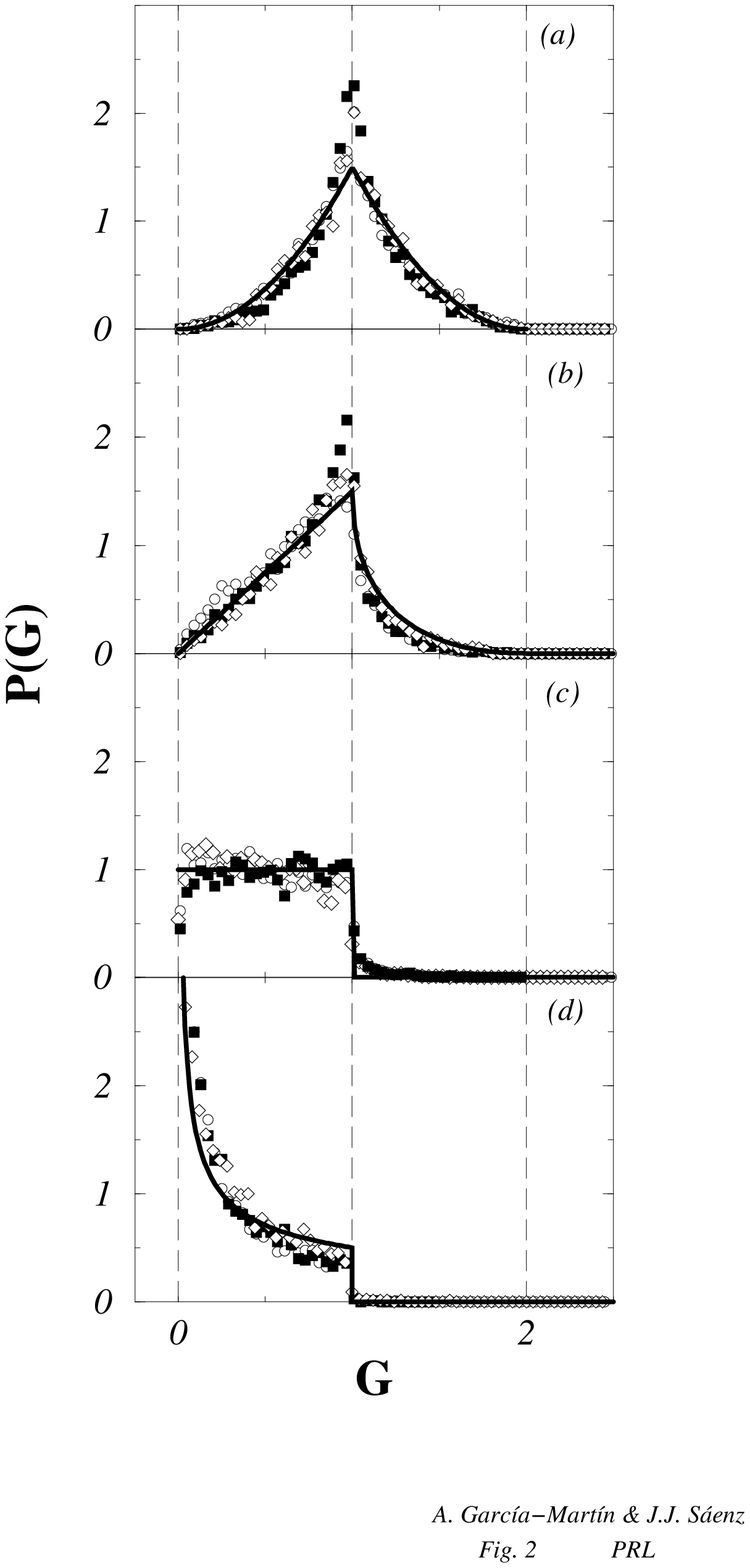,width=8.5cm,clip=}  
\caption{%
Conductance distribution for the different 
ave\-raged conductance values [{\em (a)}$\langle G\rangle=1$; 
{\em (b)}$\langle G\rangle=4/5$;{\em (c)}$\langle G\rangle=1/2$;
{\em (d)}$\langle G\rangle=1/3$]
shown in Fig. \ref{System}.  Symbols are for
$N=9$ ($\Diamond$), $N=5$ ({\large $\circ$}) and $N=3$ ($\blacksquare$).
Continuous lines represent the analytical
results of our random matrix approach.
} 
\label{Latira} 
\end{center} 

\end{figure}

The analysis of the conductance distributions $P(G)$ along the
transition from the diffusive regime $L<\xi$ to the localization regime 
$L>\xi$ (see Fig. \ref{Latira}) shows that the distributions are far from
evolve smoothly from the Gaussian to the lognormal
distribution as one could expect from the behavior of 
the transmission coefficients \cite{Lagen,PRL1}.
Our results for $P(G)$ are shown in Figure  \ref{Latira} for some selected
values of $\langle G \rangle$ around the
transition from the diffusive to the localization regime 
($\langle G \rangle \approx $ 1 (a), 4/5 (b), 1/2 (c) and 1/3 (d); 
those values correspond to the symbols in Fig. 1). The results for 
$N=9$ ($\Diamond$), $N=5$ ({\large $\circ$}) and $N=3$ ($\blacksquare$) are
almost  identical (within our
numerical accuracy) suggesting that the only parameter determining the
shape of $P(G)$ is the average conductance $\langle G \rangle$ as expected from scaling
arguments \cite{itisworth}. 

It is remarkable that  the almost perfectly flat distribution obtained for
$\langle G \rangle=1/2$ (Fig. 2c)
is in full agreement with those obtained for 2D
systems in the quantum Hall regime 
at the critical point 
\cite{Wang,Jovanovic}: 
a flat distribution
except for
a small dip at $G \approx 0$ \cite{Wang,Shapiro} and an exponential 
cutoff for $G>1$. 
As the averaged conductance goes slightly below 1/2, $P(G)$ becomes a
``one-sided'' log-normal distribution with a sharp cut off  at 
$\langle G \rangle >1$ (Fig. 3)
 in agreement with both numerical \cite{Plerou,Ruhl} and
analytical \cite{Muttalib} results. In Figure 3 we have plotted our results
for $P(\ln G)$
for different values of $\langle G \rangle$ together with the best fit to a
one-sided log-normal distribution (thick solid lines). Although the
agreement is very good for all values of $\langle G \rangle < 1/2$, the
distribution at the critical value of 1/2 is clearly better described by 
$P(\ln G, \langle G \rangle =1/2) = \exp(\ln G)$ (thin line in Fig. 3a) 
which corresponds to a
flat distribution ($P(G, \langle G \rangle =1/2) = 1$ , $0\leq G\leq 1$).
Although there is no phase transition in 
quasi-one-dimensional systems (1D), our results suggest a special generic
behavior at a critical $\langle G \rangle_c = 1/2$ which is certainly not 
consistent with a one-sided log-normal distribution.

In order to get more insight on
the exhibited behavior of the conductance
distributions we have made an extensive analysis in terms of the
transmission eigenchannels $\{ \tau_i \}$. 
These eigenchannels are the eigenvectors of
the matrix $tt^{\dagger}$ and they form the natural basis to analyze the
properties of the conductance ($G={\text{trace}}(tt^{\dagger})= \sum_i
\tau_i$). 
Following 
Imry's work\cite{Imry86}, we arrange the eigenvalues
is descending order, and perform the averaging only taking those
eigenvalues of the same level into account (e.g. $\langle
\tau_1\rangle$ corresponds to the average of the highest eigenvalue of
each realization).  
In Fig. 4 we show the evolution of the average value of the eigenchannels
as a function of $\langle G \rangle$. In the diffusive regime ($\langle G
\rangle > 1$), in agreement with the behavior discussed by Imry, almost
all the eigenchannels are either fully open or closed and only a few of
them fluctuate giving rise to UCF. However, their exact behavior depends on
the particular parameters of the wire. 
Close to
the crossover to localization, when all the eigenchannels are closed
and only two (or one in the localization threshold)
of them are partially open, the average eigenchannel transmissions
(Fig. 4) as well as the conductance distributions (Fig. 2)
do not depend on the size of the wire or the defect details. In this sense,
our results suggest an universal behavior of the
conductance with $\langle G \rangle$ as the only scaling parameter.

In contrast with the insulating regime $\langle G \rangle < 1/2$, 
where the conductance is known to be well described by one-sided 
log-normal distributions \cite{Muttalib}, 
in the crossover regime $\langle G \rangle > 1/2$, where the statistics is
dominated by one or two fluctuating channels, there is no analytical
result available for the conductance distribution. As a first 
analytical approach to this problem,
given our limited knowledge of the statistical correlations between
different eigenchannels, a possible choice of the statistical ensemble is
that which maximizes the information entropy subject to the known
constrains of flux conservation and time-reversal invariance. 
In the RMT context, this leads to the circular orthogonal ensemble. 
In our case, we also know that, close to the onset of localization 
(for $1/2 \lesssim \langle G \rangle \lesssim 1$), 
only two eigenchannels actually contribute to the conductance,
independently of the initial number of channels $N$. 
From the polar decomposition of the scattering matrix \cite{Bee,BaranJal94}
it is possible to write
the joint probability distribution of $n$ transmission eigenchannels 
(with $\tau_i=0$, $i=n+1,\dots,N$)  as
\begin{equation}
P(\{\tau_i\}) \propto \prod_{i<j}^n |\tau_i-\tau_j| \prod_k^n \tau_k^{\alpha}
\label{petete}
\end{equation}
where $\alpha=(1+n)(\langle G \rangle - n/2)/(n-\langle G \rangle)$
\cite{refpetete}.
Within this approach, the conductance distribution is fully described 
by the number of
fluctuating channels $n$ and the mean value of the conductance
$\langle G\rangle$ \cite{averages}. Simple closed expressions for
$P(G;\langle G\rangle,n)$ can be obtained for $n=1,2$,
\begin{equation}
 P(G;\langle G\rangle,1) \propto
G^{-\frac{1-2\langle G\rangle}{1-\langle G\rangle}} ,\text{ for $0 < G < 1$},
\end{equation}
and  
\begin{equation}
P(G,\langle G\rangle,2) \propto
\begin{cases}
 \left(\frac G2\right)^{\frac{2(1-2\langle G\rangle)}{\langle G\rangle-2}} 
,& \text{if $0 < G \leq 1 $} \\
 \left(\frac G2\right)^{\frac{2(1-2\langle G\rangle)}{\langle G\rangle-2}}
- \left(G - 1\right)^{\frac{1-2\langle G\rangle}{\langle G\rangle-2}},&
\text{if $1 \leq G < 2 $}
\end{cases}
\label{PG_CC2b1}
\end{equation}
In Fig. \ref{Latira} we have plotted the RMT distributions (continuous
thick line) for different $\langle G\rangle$ ($n=1$ for $\langle G\rangle
\leq 1/2$ and $n=2$ for  $1/2 \leq \langle G\rangle$),
together with our numerical results. 
As it can be seen, there is a very good agreement  
between the analytical results and the numerical calculations.
In particular, our RMT approach captures some important features of the
distributions such as the almost perfectly flat distribution at
$\langle G \rangle=1/2$ 
and the cusp point in $P(G,\langle G \rangle)$ at 
$G =1$.
Since the RMT results are quite general, we expect qualitative similar
conductance distributions in higher dimensions close to the critical
regime.

In summary, we have analyzed  the evolution of the
conductance distribution all the way from the diffusive to the
localization regime. Close to the crossover regime we have shown that 
the distributions are independent of the details of the system with 
the averaged conductance  $\langle G\rangle$ as the only scaling parameter. 
For $\langle G\rangle$ between 1/2 and 1, the statistics is dominated by 
one or two fluctuating eigenchannels and the
numerical conductance distributions are surprisingly well described by 
RMT results 
At  a critical value $\langle G\rangle_c = 1/2$, the
distribution is almost perfectly flat.
In the insulating regime ($\langle G\rangle < 1/2$) our
results are consistent with one-sided log-normal distributions.
The similarities between our results and those obtained in
very different situations suggest that 
the conductance distributions
exhibit a {\em universal behavior} at the crossover regime.

We specially acknowledge   P.A. Mello, M. Nieto-Vesperinas,  J.-L.
Pichard  and P. W\"{o}lfle for their  fruitful and stimulating suggestions. 
We are grateful
to C. Soukoulis for discussions and bringing the results of the quantum
Hall transitions to our attention. We also like to thank  A. Cano and L.S.
Froufe for interesting discussions.  This work has been supported by the
Comunidad Aut\'{o}noma de Madrid and the DGICyT through Grants 
07T/0024/1998 and  No. PB98-0464.

\begin{figure}
 
 
\narrowtext 
\begin{center} 
\epsfig{file=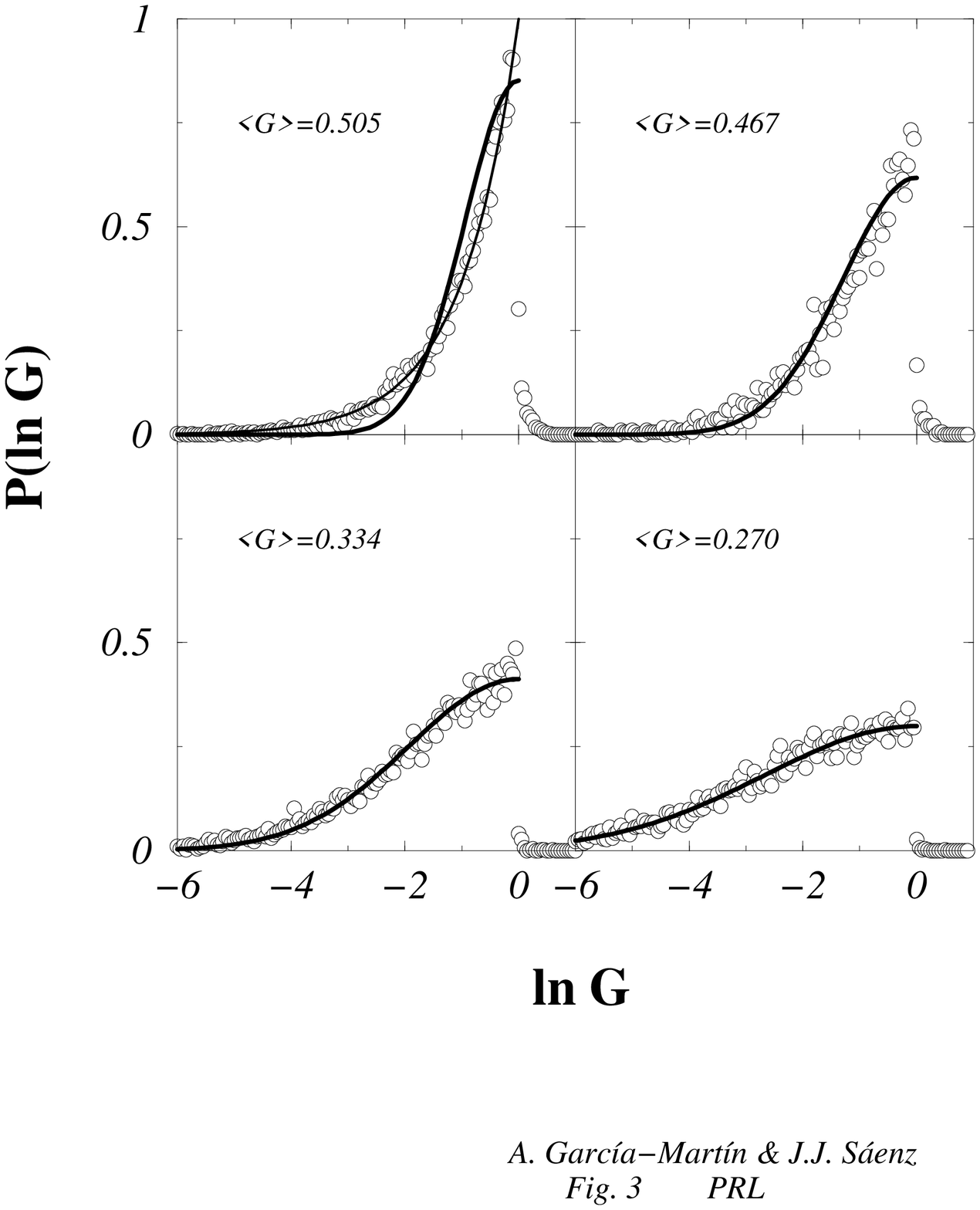,width=8.5cm,clip=} 
\caption{ 
Distribution of $\ln G$ for different values of the average conductance
beyond the critical point ($\langle G_c \rangle \lesssim 1/2$). Thick solid
lines are the best fits to a one-sided log-normal distribution. Thin solid
line in (a) corresponds to $P(\ln G, \langle G \rangle =1/2) = \exp(\ln G)$
(i.e. $P(G,\langle G \rangle =1/2)$ is a uniform distribution).}
\label{PG_2fluc} 
\end{center} 

\end{figure}

\begin{figure}
\narrowtext 
\begin{center} 
\epsfig{file=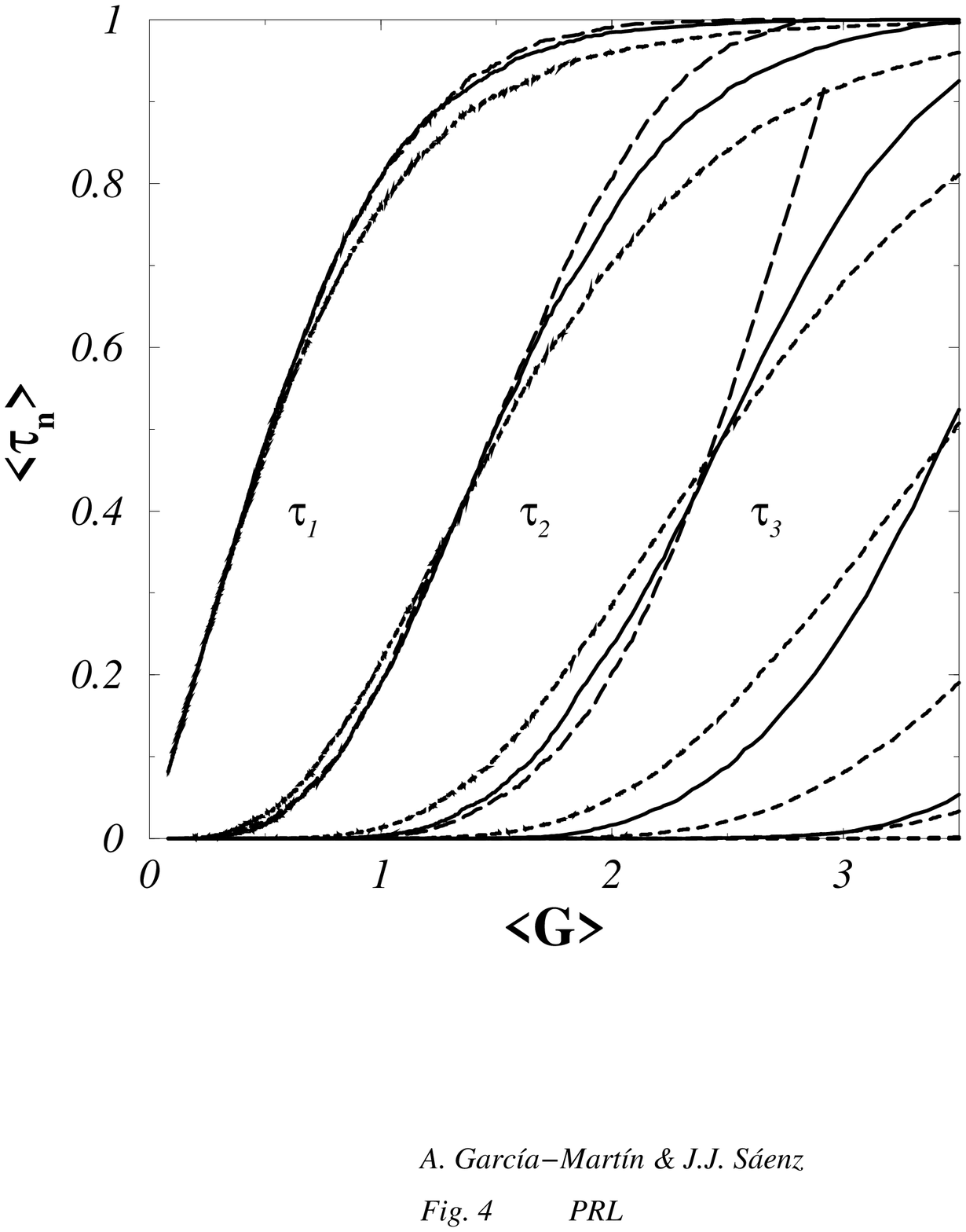,width=8.5cm,clip=}  
\caption{ 
Averaged transmission eigenvalues $\langle \tau_i \rangle$
as a function of the  averaged conductance. Dashed line corresponds to
a 9-mode wire, solid line to a 5-mode wire and long-dashed line to a
3-mode wire.} 
\label{Eigen_L_G} 
\end{center} 

\end{figure}

\end{multicols}

\begin{references}
\bibitem{averages}
In terms of the number of channels $N$, the average conductance is given by
$\langle G \rangle = n (2N-n)/(2N+1)$.
Following the results of ref. \cite{PRL2}, we consider the
distributions as a function of the average conductance rather than in terms
of the total number of channels.
\end{references}
\end{document}